\documentclass[fleqn,usenatbib]{mnras}

\usepackage{newtxtext,newtxmath}

\usepackage[T1]{fontenc}
\usepackage{soul}

\DeclareRobustCommand{\VAN}[3]{#2}
\let\VANthebibliography\thebibliography
\def\thebibliography{\DeclareRobustCommand{\VAN}[3]{##3}\VANthebibliography}

\usepackage{graphicx}	
\usepackage{amsmath}	
\usepackage{xspace}
\usepackage[dvipsnames]{xcolor}

\newcommand{\hMpc}{ h^{-1}{\rm Mpc}}

\newcommand{\hGpcC}{ h^{-3}{\rm Gpc}^3}

\newcommand{\ihMpc}{ h\,{\rm Mpc}^{-1}}
\newcommand{\mtm}{\texttt{map2map}\xspace}
\newcommand{\dg}{\delta_{\rm g}}
\newcommand{\q}{\vec{q}}
\newcommand{\x}{\vec{x}}
\newcommand{\s}{\vec{s}}

\usepackage[normalem]{ulem}

\title[Hybrid-bias and field emulators]{Hybrid-bias and displacement emulators for field-level modelling of galaxy clustering in real and redshift space}

\author[Pellejero Iba\~nez et al.]{
Marcos Pellejero Iba\~nez$^{1,2}$\thanks{E-mail: mpelleje@roe.ac.uk (MPI)},
Raul E. Angulo$^{1,3}$\thanks{E-mail: reangulo@dipc.org (REA)},
Drew Jamieson$^{4}$,
and Yin Li$^{5,6,7}$
\\
\\$^{1}$ Donostia International Physics Centre, Paseo Manuel de Lardizabal 4, 20018 Donostia-San Sebastian, Spain.\\
$^{2}$Institute for Astronomy, University of Edinburgh, Royal Observatory, Blackford Hill, Edinburgh, EH9 3HJ , UK\\
$^{3}$IKERBASQUE, Basque Foundation for Science, 48013, Bilbao, Spain.\\
$^{4}$Max-Planck-Institut für Astrophysik, Karl-Schwarzschild-Stra$\beta$e 1, 85748 Garching, Germany.\\
$^{5}$Department of Mathematics and Theory, Peng Cheng Laboratory, Shenzhen, Guangdong 518066,
China.\\ 
$^6$Center for Computational Mathematics, Flatiron Institute, New York, New York 10010, USA.\\
$^7$Center for Computational Astrophysics, Flatiron Institute, New York, New York 10010, USA.\\
}

\date{Accepted XXX. Received YYY; in original form ZZZ}

\pubyear{2015}

\begin{document}
\label{firstpage}
\pagerange{\pageref{firstpage}--\pageref{lastpage}}
\maketitle

\begin{abstract}
Recently, hybrid bias expansions have emerged as a powerful approach to modelling the way in which galaxies are distributed in the Universe. Similarly, field-level emulators have recently become possible thanks to advances in machine learning and $N$-body simulations. In this paper we explore whether both techniques can be combined to provide a field-level model for the clustering of galaxies in real and redshift space. Specifically, here we will demonstrate that field-level emulators are able to accurately predict all the operators of a $2^{\rm nd}$-order hybrid bias expansion. The precision achieved in real and redshift space is similar to that obtained for the nonlinear matter power spectrum. This translates to roughly 1-2\% precision for the power spectrum of a BOSS and a Euclid-like galaxy sample up to $k\sim0.6\ihMpc$. Remarkably, this combined approach also delivers precise predictions for field-level galaxy statistics. Despite all these promising results, we detect several areas where further improvements are required. Therefore, this work serves as a road-map for the developments required for a more complete exploitation of upcoming large-scale structure surveys.    
\end{abstract}




\section{Introduction}

The spatial distribution of galaxies encodes valuable information about the Universe. By examining the statistical features of this distribution we can infer precise details of the composition of the Universe, its expansion history, the formation history of cosmic structure, and the underlying physical laws governing these phenomena. Numerous observational surveys are poised to generate detailed 3-dimensional maps of the galaxies distribution. These maps will provide tight constraints on cosmology and gravitational physics, and could lead to a measurement of the neutrino mass and the discovery of new physics (e.g. Euclid, \citealt{Laurejis2011} and \citealt{Euclid}, and DESI, \citealt[][]{DESI}).

Extracting the maximum amount of information from these upcoming large-scale structure (LSS) surveys is an active field of research. Traditionally, the two point statistics (power spectrum and correlation function) have been the primary basis for cosmological inference (see e.g. \citealt{Cole:2005MNRAS,Blake:2010,Beutler:2017,Pellejero-Ibanez_2017}). However, since the galaxy field and the underlying dark matter are highly non-Gaussian, a large amount of information lies in higher order statistics. For instance, constraints from the BOSS survey tighten by 15--20\% when the bispectrum is jointly analysed along with the power spectrum (see e.g. \citealt{PhilcoxIvanov2022} and \citealt{IvanovPhilcox2023}). Recently, a plethora of additional summary statistics have been proposed, including k-nearest-neighbours (kNNs, \citealt{Banerjee_2021}), cosmic voids \citep{Pisani2019,Moresco2022}, wavelet scattering transforms \citep{Scattering_2020,Valogiannis_2022a,Valogiannis_2022b}, line-correlation functions \citep{Eggmeier_2015}, and artificial intelligence algorithms \citep{SIMBIG_2023}. Ultimately, the only way to guarantee that the full amount of information is extracted from observational data is to model the data at the field level, utilising the information from every single observed resolution unit -- a topic that has recently attracted a lot of attention from the community (see e.g. \citealt{Andrijaetal2022,Stadler_2023,Porqueres2023,Boruah:2023}).

A common challenge for all analysis methods, regardless of whether they use field-level or summary statistics, is to guarantee the robustness of the corresponding theoretical model. Although numerical simulations of galaxy formation have advanced enormously over the last decade, they make assumptions and simplifications about galaxy formation physics, which might not be valid in the real Universe at the precision of LSS surveys. Empirical models (such as Halo Occupation Distribution and/or SubHalo Abundance Matching) offer a more agnostic approach, but they still rely on correctly modelling the underlying processes, which cannot be formally guaranteed. Conversely, perturbative descriptions of biased tracers offer robustness, in the sense that these should be valid regardless of the details of galaxy formation as long as certain symmetries are present. Unfortunately, perturbation theory is valid only on very large scales, which implies that a large amount of information would be discarded in data analyses.

Ideally, we would like to leverage the advantages of both the accurate $N$-body model of nonlinear gravitational clustering and the perturbative description of uncertain bias relations. This can be achieved by combining a Lagrangian-space bias expansion with displacement and velocity fields measured from $N$-body simulations. Using $N$-body to transform from Lagrangian to Eulerian coordinates extends the range of validity of the perturbative expansion down to mildly nonlinear scales. Recent work by \cite{Modi2020} and \cite{Pellejero2022} demonstrated the effectiveness of this approach in accurately modelling galaxy clustering two-point statistics down to $k=0.6\ihMpc$. These authors also employed this hybrid model in real and redshift space to construct power spectra emulators (\citealt{Zennaro2021,kokron2021,Pellejero2023}), achieving significant gains for these summary statistics. Field-level analysis, however, requires generating full 3-dimensional galaxy fields, which cannot be obtained from summary statistics emulators. 

It is important to emphasise that beyond the nonlinear scale ($k_{\rm NL}>0.25h/$Mpc at $z=0$) higher-order operators in the galaxy bias expansion are not suppressed with respect to lower-order ones. The conventional convergence criterion, which permits the truncation of the bias expansion at a finite order, loses its validity in this context \citep{Desjacques2018}. Consequently, not only can previously overlooked third order operators become significant, but any contribution from higher-order terms is not guaranteed to be suppressed beyond $k_{\rm NL}$. Therefore, the applicability of the hybrid-bias model at a fixed order on non-perturbative scales should be regarded as a heuristic representation of galaxy bias, necessitating meticulous evaluation for each specific tracer population. Similarly, the range of validity will depend not only on the tracer population but also on the summary statistics one aims to model.

Exploiting the benefits of a hybrid bias approach at the field level requires making faster predictions for the nonlinear displacement field than can be directly obtained from $N$-body simulations. Speeding up this prediction typically sacrifices accuracy, such as under $n^\mathrm{th}$ order Lagrangian Perturbation theory (LPT) with or without suitable enhancements (see \citealt{Desjacques2018} for a review), or through approximate, low resolution $N$-body solvers (see e.g. \citealt{FastPM,COLA}). Another approach is to use deep learning models to emulate the $N$-body evolution at the field level, which can greatly accelerate the prediction while retaining $N$-body accuracy on relevant scales (see e.g. \citealt{He2019,Renan2020,NECOLA,Jamieson2022}).

The basic idea behind field-level emulators is to employ convolutional neural networks to learn the mapping between a fully nonlinear displacement field and an approximation of this field predicted by a fast method. The versatility of convolutional neural networks enables a broad range of applications, including super-resolution simulation \citep{Li2021,Ni2021,Zhang2023} and error correction for approximate $N$-body solvers \citep{NECOLA,Jamieson2022}. In this paper we implement the state-of-the-art field-level emulator \mtm\footnote{https://github.com/dsjamieson/map2map\_emu/} from \citealt{Jamieson2022} in our hybrid bias expansion model. This emulator extends the capabilities of neural network models by incorporating dependence on the matter density parameter, $\Omega_{\rm m}$ through style-parameters, and leveraging the Quijote Latin hypercube simulation suite \citep{QuijoteSims} with 2,000 different sets of cosmological parameters as training data. The emulator delivers percentage accurate results for the nonlinear matter power spectrum in both real and redshift space at scales of $k\sim 0.6h{\rm Mpc}^{-1}$.

In this paper we demonstrate the accuracy of a 2nd-order hybrid bias expansion model that uses \mtm to emulate the nonlinear displacements and velocities at the field level. While we reference higher-order statistics and conduct field-level analysis, this study does not specifically investigate the hybrid-bias model performance in modelling these observables. Instead, its primary focus lies in assessing the emulator's precision in replicating higher-order operators derived from comprehensive $N$-body simulations. Note that the \mtm emulator has undergone preliminary testing against $N$-body simulations as outlined in \cite{Jamieson2022}. Here, we test the emulator on a more realistic scenario, where the dark matter field is weighted by the physical properties relevant for galaxy formation.

This opens the door to fast, efficient, and accurate generation of many realisations across a wide range of cosmology model parameters. This has the potential to greatly improve our ability to estimate covariance matrices for summary statistics, reduce the noise in summary-statistic emulators, perform reconstruction of the dark matter distribution, generate key fields for Intrinsic Alignments \citep{Himalaya2023}, and even model LSS observables at the field level.

The structure of this paper is as follows. In \S\ref{sec:methods} we recap the two main tools of this study: a hybrid 2nd-order bias model (\S\ref{sec:bias}) and the \mtm emulator (\S\ref{sec:m2m}). Then, in \S\ref{sec:results} we present the results of our combined approach, first visually and then in terms of individual operators of the bias expansion. We then focus on two galaxy samples that mimic those in the CMASS-BOSS and EUCLID galaxy surveys. We explore the power spectrum in real and redshift space, as well as its predictions for the covariance. We further analyse the ability of our approach to capture variations in the underlying cosmological parameters and study the field-level comparison of simulated and emulated fields. We conclude in \S\ref{sec:conclusions}.

\section{Methodology}
\label{sec:methods}

\subsection{Hybrid bias expansion model}
\label{sec:bias}

Galaxies populate the matter field in complicated ways that have uncertain relations to environment and formation history. This connection depends on mass accretion history, mergers, and halo properties, as well as on a large variety of astrophysical processes (star formation and stellar evolution, black hole growth, feedback processes, etc). We describe this complexity by invoking a perturbative galaxy-halo connection (see e.g. \citealt{Desjacques2018} for a review). The uncertainty is then handled by measuring and marginalising over the bias parameters introduced in the perturbative expansion. Specifically, at $2^{\rm nd}$ order in perturbation theory, the galaxy overdensity in Lagrangian coordinates, $\dg(\q)$, can be written as:

\begin{equation}
    \dg = 1 + b_1\delta + b_2 (\delta^2 - \langle \delta^2\rangle) + b_{s^2} (s^2 - \langle s^2\rangle) + b_{\nabla}\nabla^2\delta \, ,
    \label{eq:bias}
\end{equation}

\noindent where $\delta$ is the smoothed matter overdensity in Lagrangian space (i.e. the linear overdensity field), $s^2$ is the traceless part of the tidal field, $s^2=s_{ij}s^{ij}=(\partial_i\partial_j\phi - 1/3\delta^{\rm{K}}_{ij}\delta)^2$, with $\phi$ being the linear gravitational potential, $\nabla^2\phi = 4\pi G \bar{\rho}\delta$. The free coefficients $b_1$, $b_2$, $b_{s^2}$, and $b_{\nabla}$ are referred to as Lagrangian bias parameters.

To make predictions for the tracers in Eulerian space, $\x$, the galaxy overdensity needs to be advected: $\dg(\q) \rightarrow \dg(\x)$. This mapping can be carried out using Lagrangian perturbation theory, or, as in our hybrid approach, using $N$-body displacements. The hybrid approach can also be extended to redshift space, where we apply the additional velocity-dependent advection along the line-of-sight, $\dg(\x) \rightarrow \dg(\s\equiv \x + v_z(\x)/aH)$. Here $v_z$ is the line-of-sight velocity measured in simulations.

In the \texttt{BACCO} implementation of hybrid biasing \citep{Zennaro2021,Pellejero2022,Pellejero2023}, the galaxy overdensity is written as a weighted sum of two terms modelling the contribution of central and satellite galaxies:

\begin{equation}
    \dg(\s) \equiv (1-f_s)\dg(\s_c) + f_s \dg(\s_c) \ast_z \rm{exp}(-\lambda_{\rm FoG} s_z) \, ,
    \label{eq:FoG}
\end{equation}

\noindent where $f_s$ and $\lambda_{\rm FoG}$ are two additional free parameters that represent the fraction of satellite galaxies and their typical velocity dispersion, respectively (FoG stands for Fingers-of-God, \citealt{JohnJackson1972,SargentTurner1977}). The star operation in the second term is a 1D convolution along the line-of-sight. $\s_c = \x + v_z^c(\x)/aH$ where $v_z^c$ is the line-of-sight velocity of the parent halo. This convolution is performed on all regions of Lagrangian space that are found inside collapsed structures. The main motivation for this is that we expect central and satellite galaxies to have very different small-scale velocities: central galaxies are predominantly at rest with respect to their halo whereas satellite galaxies have considerable random motions, giving rise to the so-called Finger-of-God effect. The magnitude of this effect scales with the virial velocity of the parent halo, but in detail depends on galaxy formation physics and the galaxy types that are sampled \citep[see][for an extended discussion]{OrsiAngulo2018}. Additionally, we consider a stochastic term $( \epsilon_1+\epsilon_2k^2)/\bar{n}_{\rm tr}$ which parameterizes unmodelled small-scale physics and terms that are not included in our bias expansion \citep[e.g.][]{Perko2016}. Here, $\epsilon_1$ and $\epsilon_2$ are two free parameters, and $k$ stands for the Fourier space wavenumber.

Note that Eq.~\ref{eq:bias} is an approximation to the relationship between galaxies and dark matter, which is strictly valid only on large-scales, $\delta < 1$. As we consider smaller scales, $3^{\rm rd}$, higher-order bias terms that we have neglected can become important, potentially affecting cosmological inferences. For these reasons, the accuracy of a hybrid bias expansion needs to be monitored carefully: by checking the consistency of results with the range of scales  and the order of the expansion considered and by contrasting the model against state-of-the-art galaxy formation models.

By performing the advection of Lagrangian operators with $N$-body displacement fields, the hybrid bias model significantly improve the accuracy of the bias expansion \citep{Modi2020,Pellejero2022,DeRose2023}. This is because the predictions are no longer limited by the inaccuracies of perturbation theory in predicting nonlinear displacement and velocity fields. Similarly, this approach accounts for the nonlinearity of the transformation between the Lagrangian and redshift-space coordinates \citep{Pellejero2022}.

Previous work demonstrated that the hybrid model delivers unbiased results for the power spectrum down to scales of $k\sim0.6\ihMpc$ when compared against various galaxy formation models. Specifically, \cite{Zennaro2022} showed that the hybrid model describes the galaxy auto power spectrum and its cross-correlation with matter for 8000 different catalogues spanning various cosmological models, redshifts, selection criteria, and number densities. These tests were carried out against thousands of parameter sets for SubHalo Abundance Matching extended (SHAMe, \citealt{ContrerasAnguloZennaro2020AB, ContrerasAnguloZennaro2020}) -- an empirical extension of subhalo abundance matching that can closely resemble the hydrodynamical simulation MilleniumTNG \citep{MilleniumTNG2022}, as well as various semi-analytic galaxy formation catalogues. Additionally, \cite{Pellejero2022} showed that the hybrid model can reproduce the monopole, quadrupole, and hexadecapole of the power spectrum of SHAMe galaxies in redshift space while recovering unbiased cosmological parameters \citep{Pellejero2023}.

\subsection{The \texttt{map2map} field-level emulator}
\label{sec:m2m}

A key ingredient of hybrid models is the displacement and velocity fields computed using $N$-body simulations. Unfortunately, carrying out high-resolution simulations is computationally expensive, which limits the applicability of such models. Therefore, hybrid models were previously only built for summary statistics (such as the power spectrum or kNNs, see e.g. \citealt{Zennaro2021,kokron2021,Yuan2023}). While these will be valuable tools for analysing forthcoming LSS observations, they fall short of fully exploiting the method and the data.

The emergence of field-level emulators offers an ideal companion to hybrid biasing. These emulators provide computationally efficient methods to generate displacement fields with accuracy comparable to $N$-body simulations over the range of scales where hybrid biasing is applicable. In this work we employ the latest incarnation of \mtm field emulator, presented by \cite{Jamieson2022}. \mtm employs convolutional Neural networks (CNNs) to predict the nonlinear displacement and velocity fields of a given region of the Universe. These CNNs receive the $1^{\rm st}$-order LPT displacement field together with the value of the matter density parameter $\Omega_{\rm m}$ as inputs.

Specifically, \,\mtm uses two CNNs with a 4-tier V-Net architecture. The loss function for the displacement V-Net was set as a linear combination of the logarithm of the mean squared error ($\log$-MSE) in the Lagrangian displacements of particles and the Eulerian density field. For the velocity V-Net, the loss function was set as the sum of the $\log$-MSE in the particle velocities, the Eulerian momentum field, and the second moment of the Eulerian momentum field.

The emulator was trained using $\sim$1700 $N$-body simulations of the Quijote suite \citep{QuijoteSims}, which have distinct cosmological parameters sampled on a Latin hypercube over the 5-dimensional parameter space: 
\begin{eqnarray}
    \Omega_{\rm m} &\in [0.1 , 0.5], \\
    \Omega_{\rm b} &\in [0.03 , 0.07], \\ 
    h &\in [0.5 , 0.9], \\
    n_{\rm s} &\in [0.8 , 1.2], \\ 
    \sigma_8 &\in [0.6 , 1.0].
\end{eqnarray}
Each of these simulations evolved $512^3$ particles over a $1\,\hGpcC$ volume, which corresponds to a particle mass resolution range $m_\mathrm{p}\simeq 2\times10^{11}$ -- $1\times10^{12}~M_{\odot}\,h^{-1}$.

\subsection{The emulated galaxy bias field}
\label{sec:gals}

To validate our approach, we build two biased fields that resemble galaxy samples typically targeted by LSS observational surveys.

\begin{itemize}
    \item {\bf LRGs:} The first set mimics massive galaxies, such as those targeted by the BOSS-CMASS samples. Specifically, we employ a set of Lagrangian bias parameters $\{b_1, b_2, b_{s^2}, b_{\nabla}, \lambda_{\rm FoG}, f_{\rm sat}\} = \{0.45,0.15,0.52,-1.8,0.4,0.2\}$, and a number density of $\bar{n} = 4\times 10^{-4}/D^2(z=0.61)=2.3\times 10^{-4}[\rm{Mpc}/h]^{-3}$. We refer to this sample as Luminous Red Galaxies, LRGs.

    \item {\bf ELGs:} The second set aims to resemble emission line galaxies such as those to be targeted by the DESI and EUCLID surveys. We employ $\{b_1, b_2, b_{s^2}, b_{\nabla}, \lambda_{\rm FoG}, f_{\rm sat}\} = \{0.38,-0.36,1.44,-1.35, 0.4,0.2\}$ with a density of $\bar{n} = 6\times 10^{-4}/D^2(z=1)=2.2\times 10^{-4}[\rm{Mpc}/h]^{-3}$. We refer to this sample as Emission Line Galaxies, ELGs.
\end{itemize}

Note that the value of the bias parameters was set so that the resulting galaxy power spectrum closely matches that of physically-motivated galaxy formation models. Specifically, for the LRG's we choose the values provided by \cite{Pellejero2022} that best fit the clustering of stellar-mass selected galaxies as predicted by the SHAMe model with parameters that reproduce results in the TNG300 hydrodynamical simulation. For the ELG sample, we used the best-fit parameters of the Euclid mocks Model 3, as explained in \cite{Pozzetti2016}. In addition, since the \mtm emulator has only been trained at $z=0$, we use this redshift for our analysis but tune the value of the discreteness noise so that the ratio $P(k=0.2)/\bar{n}$ is the same between our $z=0$ catalogue and that expected for LRG and ELG samples at $z=0.61$ and $z=1$, respectively. When computing the 2-point statistics of the simulated galaxy bias fields, we further set the stochastic parameters $\{ \epsilon_1,\epsilon_2\}$ to zero.

Operationally, we start by defining a specific realisation of the initial Gaussian field on a $V=1\hGpcC$ using $512^3$ grid points, which we then smooth on a scale $k_{\rm d}=0.75\ihMpc$. We then build all 5 operators, $\{1, \delta, \delta^2, s^2, \nabla^2 \delta\}$, advect them to Eulerian space and subsequently to redshift space if appropriate. We do this either using displacement fields from \mtm or from one of the Quijote $N$-body simulations. These advected fields are finally combined and weighted with the appropriate bias parameters to retrieve a galaxy-like biased field. 

We will quantify the accuracy of the emulator combined with the hybrid bias expansion by comparing statistics of the fields generated with \mtm to those constructed from the original Quijote simulations. Although \mtm has previous been shown to deliver accurate predictions for the nonlinear power spectrum, achieving similar performance for the biased field is not guaranteed. The bias operators weigh various parts of the cosmic web differently. This could potentially reveal shortcomings of the emulated fields that do not significantly contribute to the unbiased, dark matter power spectrum.

\subsubsection{Test suite}

We employ two different datasets to evaluate the accuracy of our approach. The first one corresponds to 100 randomly-selected realisations of the ``fiducial''-cosmology simulations of the Quijote suite: $\Omega_{\rm m}=0.3175$, $\Omega_{\rm b}=0.049$, $\sigma_8=0.834$, $h=0.6711$, and $n_{\rm s}=0.96$. These simulations have the same volume and spatial resolution employed for the training of the \mtm emulator. With this suite, we characterise the typical accuracy of \mtm in predicting galaxy statistics. The second test suite corresponds to another $100$ simulations with varied cosmological parameters. None of the chosen test simulations were employed during training, which allows us to quantify the ability of \mtm at capturing the correct statistical and cosmology dependencies.

\subsubsection{Computational performance}

In addition to the robustness against galaxy formation physics, a key advantage of our approach is its computational efficiency. Unlike other approaches, such as halo occupation distribution, we do not rely on halo group finding or high spatial resolution, which makes our approach extremely computationally efficient. 

To generate a $V=1\hGpcC$ galaxy field, our model requires approximately 2.5 minutes of CPU time. Of this, approximately 30 seconds is employed to generate an initial density field and the bias operators; 15 seconds for computing 1LPT displacements, and 1 minute for augmenting them with \mtm. A final 15 seconds is required for the advection and calculation of the full Eulerian field. In redshift space, an extra minute is required for the computation of the non-linear velocity components by \mtm, although this could be further parallelized by running the velocity model on a separate node.

\begin{figure*}
	\includegraphics[width=1.\textwidth]{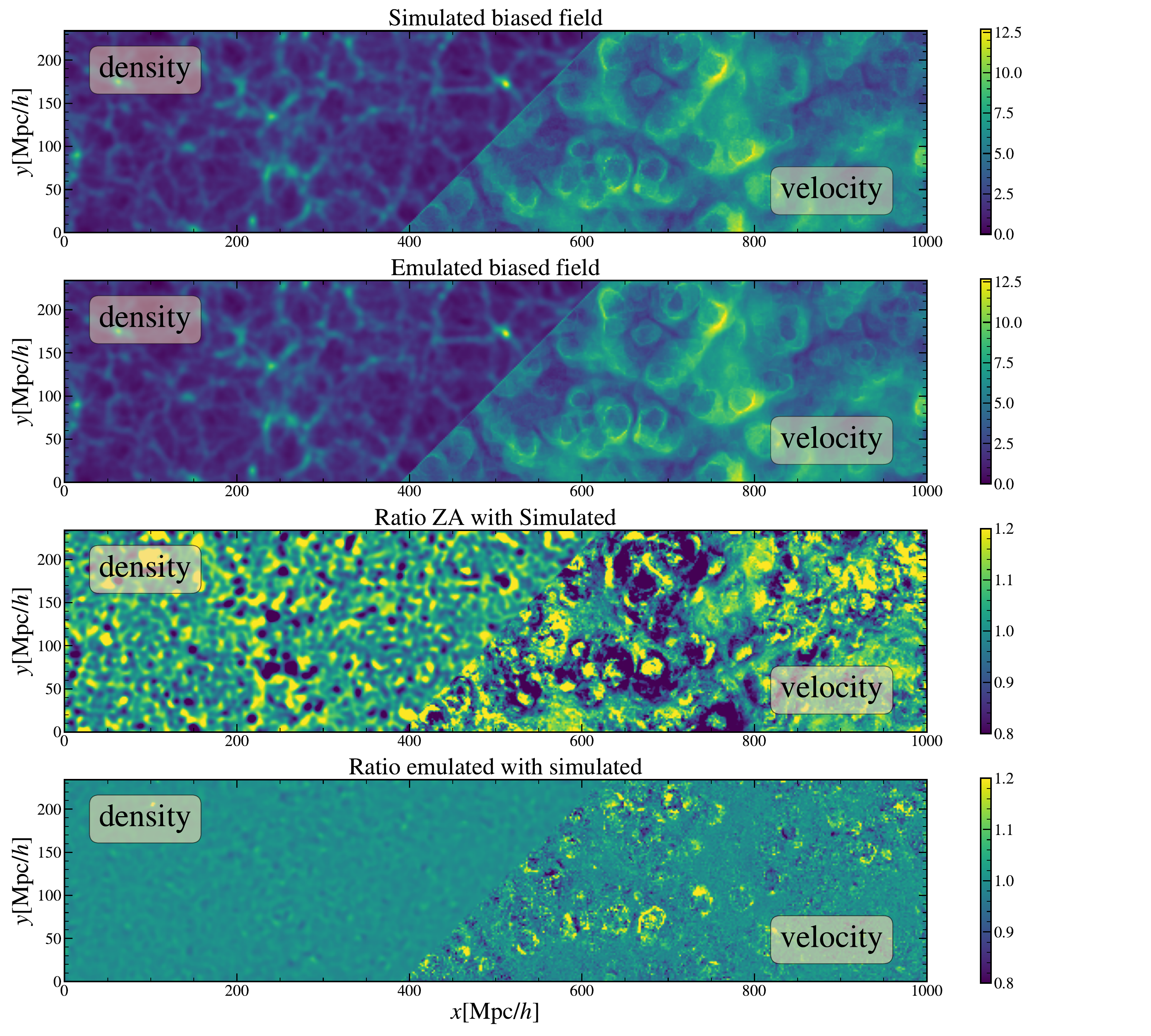}
    \caption{Comparison between biased fields computed using $N$-body simulations and the \mtm emulator. The figure shows a case study that emulates an ELG galaxy sample at $z=0$ over a spatial region of $1000\hMpc \times 250\hMpc$ with a depth of $15\hMpc$. The left and right regions of each panel depict the projected matter density and the modulus of the peculiar velocity $||\vec{v}||$, respectively. The velocity field is constructed by incorporating the velocities of the halos and the velocities of the dark matter particles not associated with any halo, as required by the hybrid model. The top panel displays the results obtained from simulated displacement and velocity fields, while the second panel showcases the corresponding outcomes using emulated quantities. For comparison, the third panel presents the ratio with respect to 1LPT. The lower panel shows the ratio between the simulated and emulated results.
    }
    \label{fig:example}
\end{figure*}

\begin{figure*}
\begin{center}	
 \includegraphics[width=1.\textwidth]{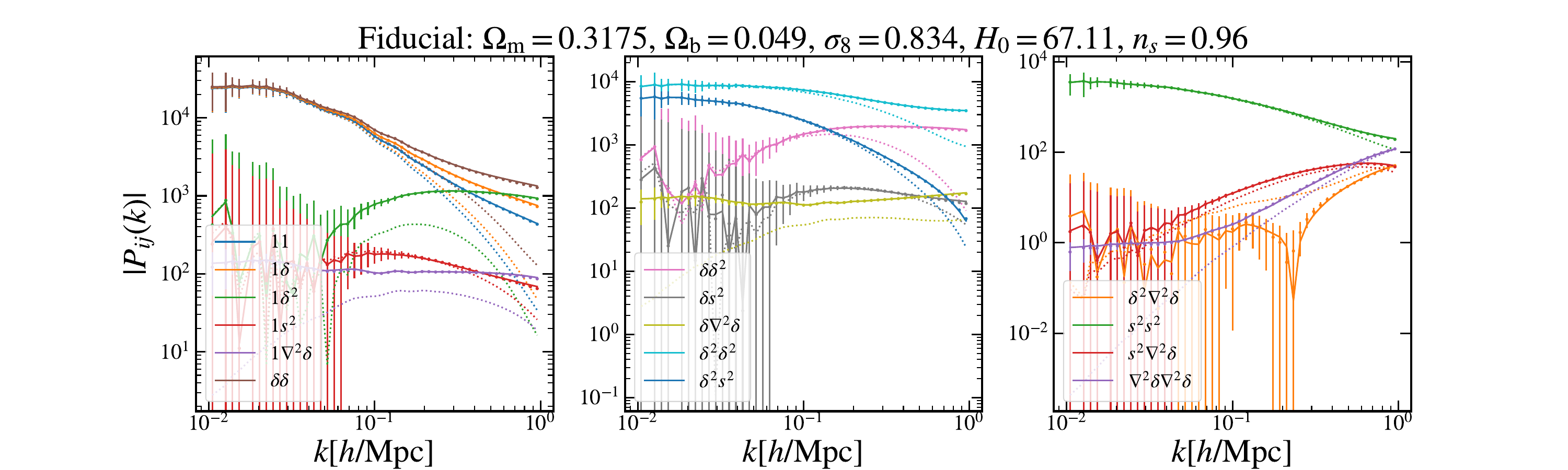}
    \caption{Power spectra of the 15 different operator fields at $z=0$ that determine a 2nd-order bias expansion, as indicated by the legend. We display our results at the fiducial cosmology of the Quijote suite, the parameter values are displayed in the figure title. For each of the power spectra, symbols display the average over 100 $N$-body realisations of a $1\hGpcC$ volume, whereas solid lines display the same quantity as predicted by \mtm. Vertical error bars indicate the standard deviation of the $N$-body ensemble. The dotted lines represent the 1LPT predictions that \mtm takes as input.}
    \label{fig:realspace1}
\end{center}	
\end{figure*}

\begin{figure*}
	\includegraphics[width=1\textwidth]{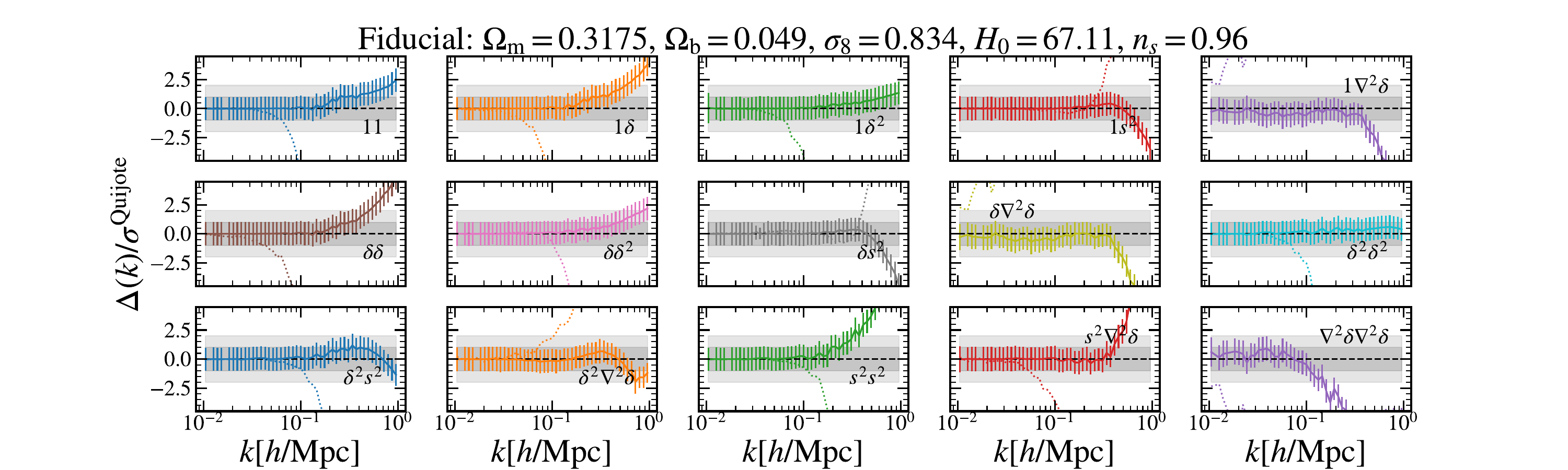}
    \caption{Relative difference between the predictions of \mtm and measurements in the Quijote suite, in units of the simulation variance for a $1\hGpcC$ volume, $(P_{\rm m2m} - P_{\rm Quijote})/\sigma^2_{\rm Quijote}$. Each panel shows results for a different pair of operator fields advected to $z=0$, as indicated by the labels on each panel. We display the average and standard deviation over 100 realisations as solid lines and vertical error bars, respectively. For comparison, dotted lines show the predictions from 1LPT displacements. Dark and light grey regions represents the 1-$\sigma$ and 2-$\sigma$ regions respectively.}
    \label{fig:realspace2}
\end{figure*}

\section{Results} 
\label{sec:results}

\subsection{Visual inspection}
\label{sec:visual_inspection}

We first illustrate the performance of our approach in Fig.~\ref{fig:example} which compares the predictions for the ELG-like galaxy sample at the fiducial cosmology using either $N$-body or emulated displacements fields. As we have argued above, the hybrid bias is expected to perform well on scales $k<0.6\ihMpc$ (at the 2-point statistics level). Therefore, we display fields after smoothing with a Gaussian filter of size $\pi/0.6 \simeq 5\hMpc$.

By comparing the top two left panels, we can appreciate a striking similarity between the emulated and simulated biased density fields. In particular, the location of voids, filaments, and overdensities coincides almost perfectly. We see that our approach is indeed capturing the non-Gaussian features of the galaxy density field. In the bottom two panels, we provide the ratio between these two predictions, we see that if displacements are provided by 1LPT (i.e. the Zeldovich Approximation, ZA), then there exist larger residuals that have a coherence length of up to hundreds of Mpc. The typical value of the residuals decreases by several orders of magnitude when \mtm displacements are used instead.

As shown by the right-side regions of Fig.~\ref{fig:example}, the accuracy of our approach is also high for the velocity field. Whereas the ZA residuals display values of up to 50\% that are coherent over tens of Mpc, the \mtm emulation greatly decreases these errors. We do find residual differences at the borders of halo structures. The origin of these discrepancies is an additional small-scale velocity dispersion in the \mtm predictions. In the \texttt{BACCO} hybrid model, Lagrangian elements that are inside haloes are assigned exactly the same velocity, given by the center of mass velocity of the parent halo. This is a challenging quantity to predict for \mtm, since it needs to correctly predict which Lagrangian elements will become part of a halo based on the displacement field. It also needs to accurately detect the boundaries of halos to distinguish the transition between particles outside and inside halos. Although not shown here, we have verified that \mtm does display an additional velocity dispersion in and around haloes. While the magnitude of this velocity is small, it is statistically significant. As we found in \cite{Lopez-Cano2023}, a tailored loss function based on Instance Segmentation techniques is required for correcting such predictions. Note that emulators trained to predict summary statistics average out such nuances, so we expect them to be much more accurate at a given statistic. There are two possible alternatives to address this issue. The first one is to improve the accuracy of \mtm. This can be achieved by either increasing the training data, improving the network architecture, or exploring an alternative formulation of the redshift-space hybrid model in terms of quantities that \mtm can predict more easily. The second alternative is to redefine the meaning of the nuisance parameters of the hybrid model so that they account for the additional velocity dispersion in \mtm. Specifically, our RSD model contains a parameter $\lambda_{\rm FoG}$ that controls the amount of velocity dispersion our modelled galaxies display inside halos. The actual value of this parameter depends on the galaxy selection. Even for a sample of purely central galaxies, we expect a small-scale velocity dispersion arising from the relative velocity between these galaxies and their host halos (an effect commonly referred to as velocity bias). Therefore, we expect $\lambda_{\rm FoG}$ to be degenerated with the intrinsic scatter of the emulation. 

In summary, this indicates a possible limitation in our approach and an avenue for potential improvement: extending the emulator's ability to predict the interior of collapsed regions in $N$-body simulations. However, as we will demonstrate below, it is possible to absorb some of these residual differences by incorporating them into the nuisance parameters of the model, which minimises their impact on our predictions. We explore further the performance of our model next.

\subsection{Power spectrum}

In this subsection, we compare the power spectrum of biased tracers obtained from emulated and simulated displacement fields. By using two different methods to fit the four nuisance parameters ${\epsilon_1, \epsilon_2, f_{\rm sat}, \lambda_{\rm FoG}}$ in the model, we obtain two distinct sets of results. The first set (named \textit{emulated}) is comprised of outcomes where the nuisance parameters are fixed from the $N$-body data, and prevented from absorbing any errors arising from the emulated displacements. The second set (named \textit{calibrated}) incorporates the ability of the nuisance parameters to accommodate any inaccuracies in the emulator. Practically, the calibration procedure is performed by minimising the distance between the emulated data and the simulated data weighted by the expected error at each wavemode. Note that both \textit{emulated} and \textit{calibrated} results are obtained from the \mtm emulated displacements. 

\subsubsection{2nd-order operators}

As explained earlier, any galaxy field is described by a sum of 5 fields weighted by their respective bias parameters under the 2nd-order hybrid bias expansion. This implies that any galaxy power spectrum will be described as a linear combination of 15 cross-power spectra.

We show these power spectra in Fig.~\ref{fig:realspace1} at $z=0$ for the fiducial cosmology of the Quijote suite. Each panel shows a distinct set of spectra. The symbols correspond to results obtained from fields advected using the the $N$-body simulation displacements averaged over 100 realizations, with error bars depicting the standard deviation among these. Solid lines show the corresponding measurement but using \mtm emulated displacements. For comparison, the spectra computed using 1LPT displacements are shown as dotted lines.

Overall, we see excellent agreement for all 15 spectra over the full range of scales considered. The \mtm emulator is able to correct for the non-linearities missing in the 1LPT approach. Specifically, the $P_{11}$, $P_{1\delta}$, and $P_{\delta\delta}$ terms -- which dominate on large scales -- are all almost indistinguishable between the simulation and emulation results. On smaller scales, quadratic terms ($P_{\delta^2 \delta^2}$ and $P_{\delta \delta^2}$) become increasingly important. These are also extremely well reproduced.  

We can better appreciate the accuracy of \mtm in Fig.~\ref{fig:realspace2}, which shows the fractional difference among each of these cross spectra. For clarity, we have displayed the differences relative to the standard deviation measured among the Quijote realisations. As in the previous figure, the dotted lines depict results from 1LPT. The error bars correspond to the 1-$\sigma$ scatter measured among the \mtm spectra. Therefore, the closer this scatter is to 1, the better the diagonal terms of the covariance matrix are recovered. We will explore this further in a subsequent section. 

On large scales, $k < 0.1 \ihMpc$, the emulated and simulated cross-spectra are in good agreement. The 1LPT predictions, on the other hand, deviate from our simulation results on scales $k \sim 0.05\hMpc$. On smaller scales, we detect systematic deviations -- depending on the spectra, $\mtm$ can overpredict or underpredict the signal by up to $3\sigma$. Note that the magnitude of this bias is nevertheless extremely small. For the $P_{11}$ term this error corresponds to a bias of 1.5\% at $k=1\ihMpc$. We present the corresponding redshift space results in Appendix \ref{App:errors}. The redshift space results include the central velocity ($v_z^c$) distortion effect, which differs between \mtm and Quijote. This leads to enhanced errors compared to the real space operators. We can contrast these findings with our earlier emulators outlined in \cite{Zennaro2021} and \cite{Pellejero2023}. It's important to note that these emulators were constructed solely based on power spectra using fully connected neural networks. They do not address the nonlinearities inherent in the $N$-body particle evolution but rather focus on their impact on two-point statistics. As indicated by \cite{Zennaro2021}, the precision of such emulators achieves values below 2\% for the scales examined in this study. This aligns with the outcomes observed for the "11" term. However, the remaining terms show values around $\sim$2-3\%, typically $\sim$0.5\% less accurate than the "direct" emulators. In redshift space, a comparison can be drawn with our prior emulator from \cite{Pellejero2023}. Once again, the accuracy hovers around $\sim$2\% (1\% for the monopole and 3\% for the hexadecapole), while \mtm attains an accuracy of approximately 3\% for the Quijote simulations. This worse performance in redshift space is attributed to the inability of \mtm to recover the halo scales accurately. Note that the absolute and relative importance of the uncertainty in each $P_{i,j}$ does, however, depend on the specific set of bias parameters describing a given galaxy sample, which is what we investigate next.

\subsubsection{Real space galaxy mock}
\label{sec:pk_real}

\begin{figure*}	
    \includegraphics[width=\textwidth]{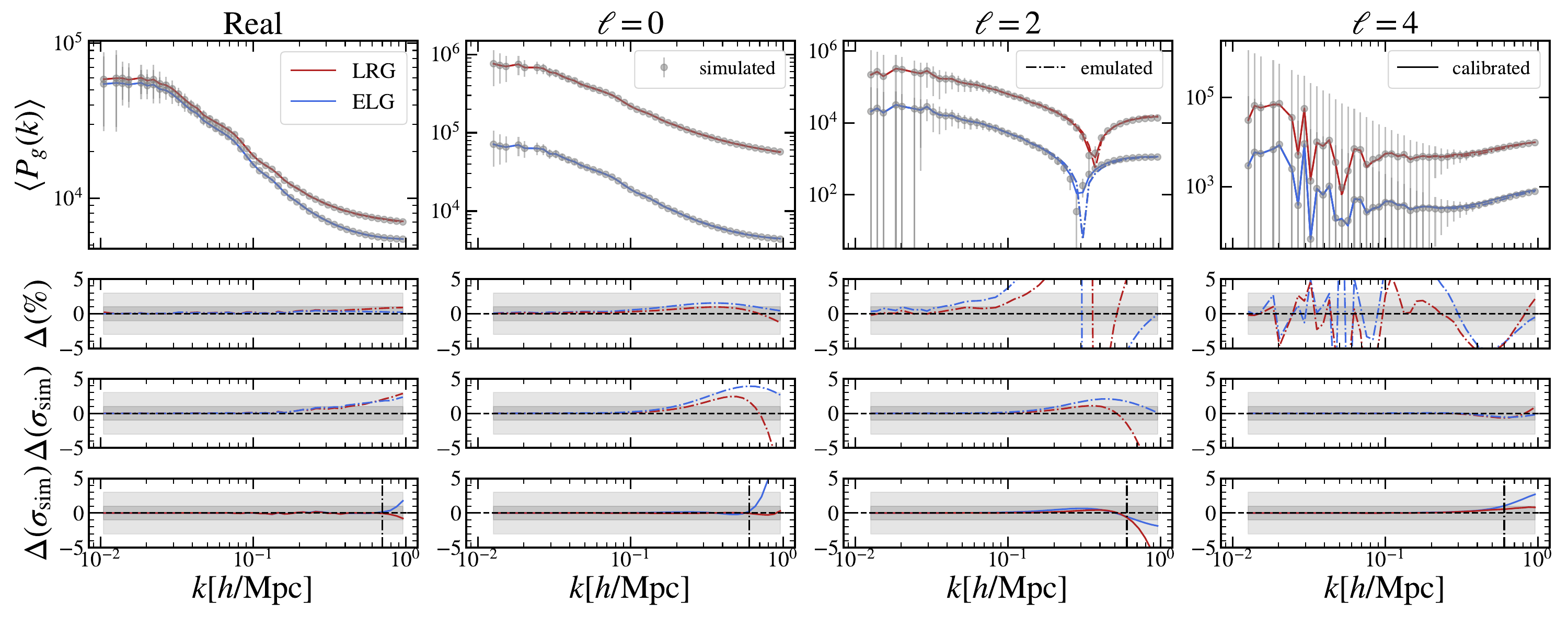}
    \caption{Comparison between the power spectra of LRG and ELG biased fields employing either Quijote simulations or the \mtm field emulator. LRG results have been displaced by one order of magnitude to clearly distinguish them. We show the results for the mean of 100 realisations in a $1\hGpcC$ volume. Bottom panels display the relative differences in various manners: first showing the percentage difference and then in units of the variance measured in the Quijote suite. In the bottom panel, we also show the case where the noise and FoG parameters in emulated galaxy fields have been calibrated to match the simulated results. Dark and light grey regions represents the 1-$\sigma$ and 3-$\sigma$ regions respectively (bottom panels) and the 1\% and 3\% regions (upper panels).}
    \label{fig:redshiftspacegalpower}
\end{figure*}

We compare the real-space power spectra of biased fields in the leftmost column of Fig.~\ref{fig:redshiftspacegalpower}. LRG and ELG tracers are displayed by red and blue lines, respectively. In the top panel, we show the mean of our 100 test realisations. The expected scale of validity of the hybrid approach for the two-point statistics, $k \sim 0.6\hMpc$, is marked by a vertical dot-dashed line in the lowest panel.
	
The \mtm results are almost indistinguishable from those obtained from the simulations, which quantitatively confirms our previous findings. As we show in the bottom panels of Fig.~\ref{fig:redshiftspacegalpower}, the fractional differences are subpercent on all scales for ELGs and barely 1\% at $k\sim1\hMpc$ for the LRG sample. These differences, however, are statistically significant given the volume of our simulations ($V=1\hGpcC$), reaching $2\sigma$ at $k\sim0.6\hMpc$. Despite this, we show that these small differences can be absorbed by the free parameters of the hybrid bias model. Specifically, in the bottom panel we adjust the value of the free parameters $\epsilon_1$ and $\epsilon_2$, which are typically interpreted as describing stochastic noise and truncated terms in the hybrid bias model. The differences are now almost absent, meaning the errors introduced by the emulator are predominantly degenerate with the errors associated with the approximations of the hybrid model itself. After freely varying the nuisance parameters, the excess error associated with the emulator is negligible.

\subsubsection{Redshift space galaxy mock}
\label{sec:pk_refshift}

A more challenging test is to reproduce the multipoles of the redshift-space power spectrum. The emulator not only needs to correctly displace Lagrangian elements to Eulerian space, it also needs to assign their correct nonlinear velocity. For our model, this task also requires knowing whether a particle is part of a halo or not.  

We show our results in the right panels of Fig.~\ref{fig:redshiftspacegalpower}. Second, third, and fourth panels show the results for the monopole ($\ell = 0$), quadrupole ($\ell = 2$), and hexadecapole ($\ell = 4$), of the power spectrum. As in previous figures, LRG and ELG galaxies are shown with red and blue lines, respectively. For clarity, we have vertically displaced the LRG results by one order of magnitude. Bottom panels show both the percentage relative differences and the differences relative to the variance as measured in our test suite.

We see that the monopole is recovered better than 2\% for both galaxy samples. The errors from both emulated galaxy samples, however, display a bump on small scales, which can be of around $5\sigma$ in magnitude at $k=0.6\hMpc$. The quadrupole shows a consistent picture: large scales are accurately recovered by \mtm but small scales are enhanced and then washed out relative to the simulated galaxy fields. The associated errors can be up to $3\sigma$ at $k\sim0.6\hMpc$. The hexadecapole is less sensitive to these inaccuracies and \mtm is statistically unbiased over almost all the range of scales we explored. The origin of these discrepancies is the additional small-scale velocity dispersion in the \mtm predictions discussed in Sec.~\ref{sec:visual_inspection}. 

In the bottom-most panel of Fig.~\ref{fig:redshiftspacegalpower} we compare our simulated galaxies with \mtm after fitting the values of our noise and RSD nuisance parameters $\{\epsilon_1, \epsilon_2, f_{\rm sat}, \lambda_{\rm FoG}\}$. As we had anticipated, we see that the freedom in our model is sufficient to absorb the effect of \mtm inaccuracies, delivering power spectrum multipoles that are almost indistinguishable from the simulated results. When considering the monopole, the primary dependencies are absorbed by the parameters $\epsilon_1$ and $\epsilon_2$. In contrast, for the quadrupole and hexadecapole, these dependencies are absorbed by the parameters $f_{\rm sat}$ and $\lambda_{\rm FoG}$, which effectively account for the influence of \mtm intrinsic scatter.

To place our results in context, we consider the case of the Euclid survey. Euclid is expected to observe 15,000 sq deg in bins of $\Delta z \sim 0.2$, which corresponds to approximately $10 \hGpcC$ -- 10 times the volume of each Quijote simulation. Even for this case, where the precision of the measurement would be approximately 3 times better than in our tests, we expect the accuracy of \mtm to affect the analysis only at scales where $k \gtrsim 0.3 \ihMpc$. Note that we could also calibrate our free parameters for lower $k$-values than $k \sim 0.6 \ihMpc$ to alleviate the emulator errors here.

\subsubsection{Variance}
\label{sec:variance}


\begin{figure*}	
\includegraphics[width=1.2\textwidth]{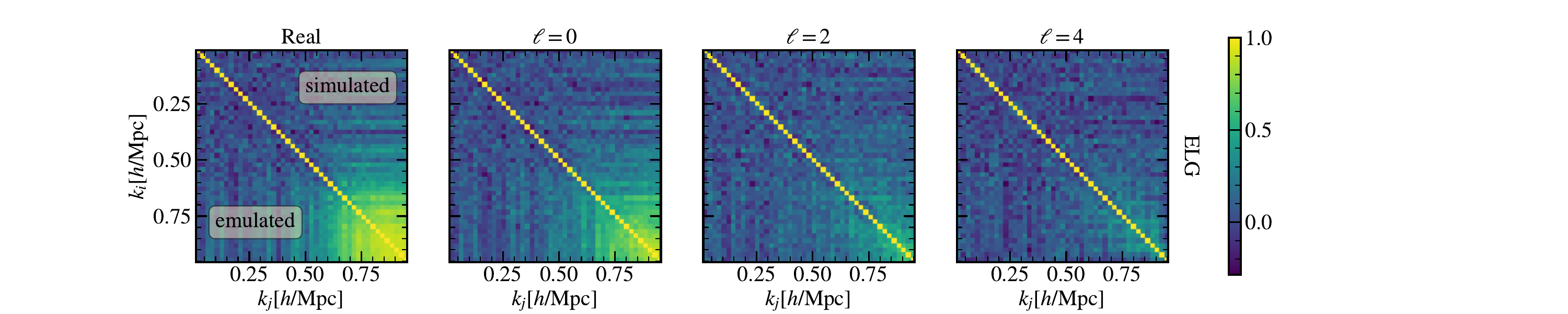}
\includegraphics[width=1.2\textwidth]{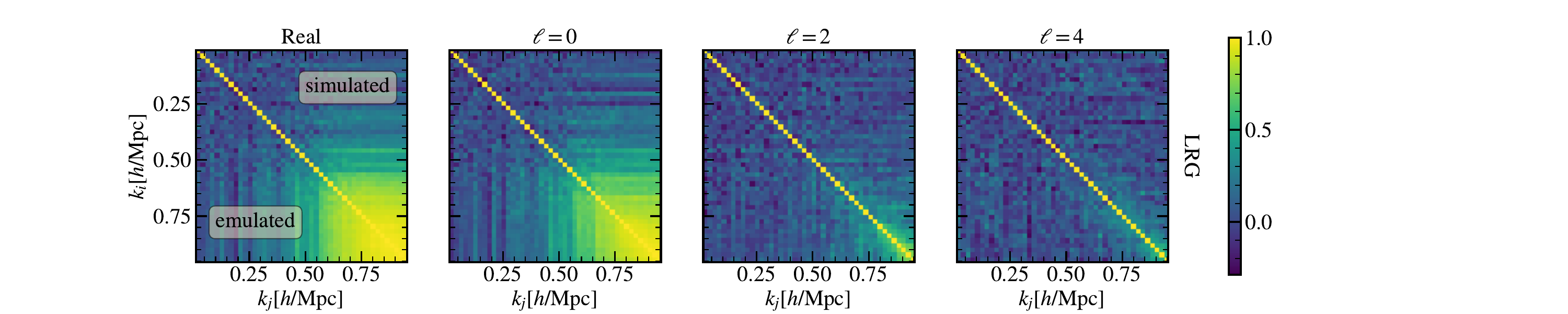}
\caption{ Correlation matrices for the ELG and LRG samples. The correlation coefficient, defined as $r(k_i,k_j)=C_{ij}/\sqrt{C_{ii}C_{jj}}$, is used to measure the correlation between different wavenumbers $k_i$ and $k_j$. The lower triangles of the matrices illustrate the results obtained from the \mtm emulator, while the upper triangles display the corresponding outcomes from the Quijote simulations conducted under the fiducial cosmology.
}
    \label{fig:covariances}
\end{figure*}

\begin{figure*}	
\includegraphics[width=\textwidth]{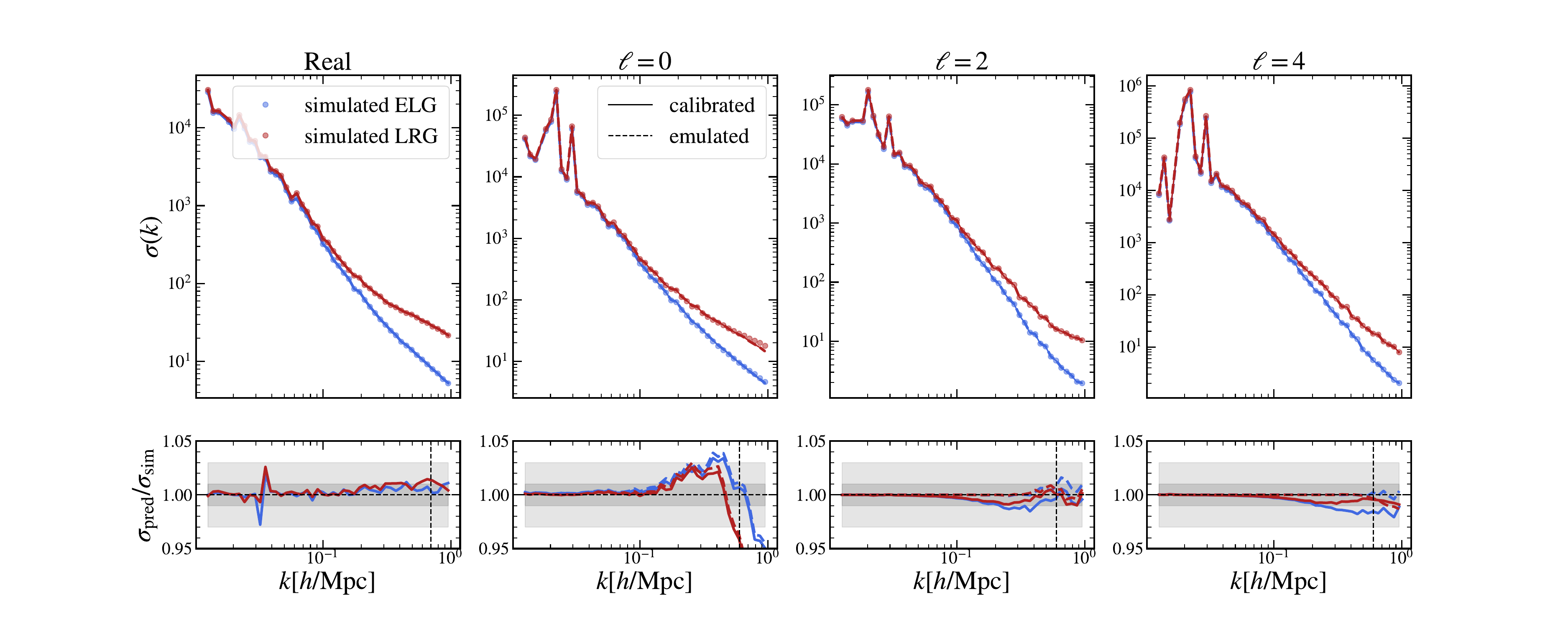}
\includegraphics[width=\textwidth]{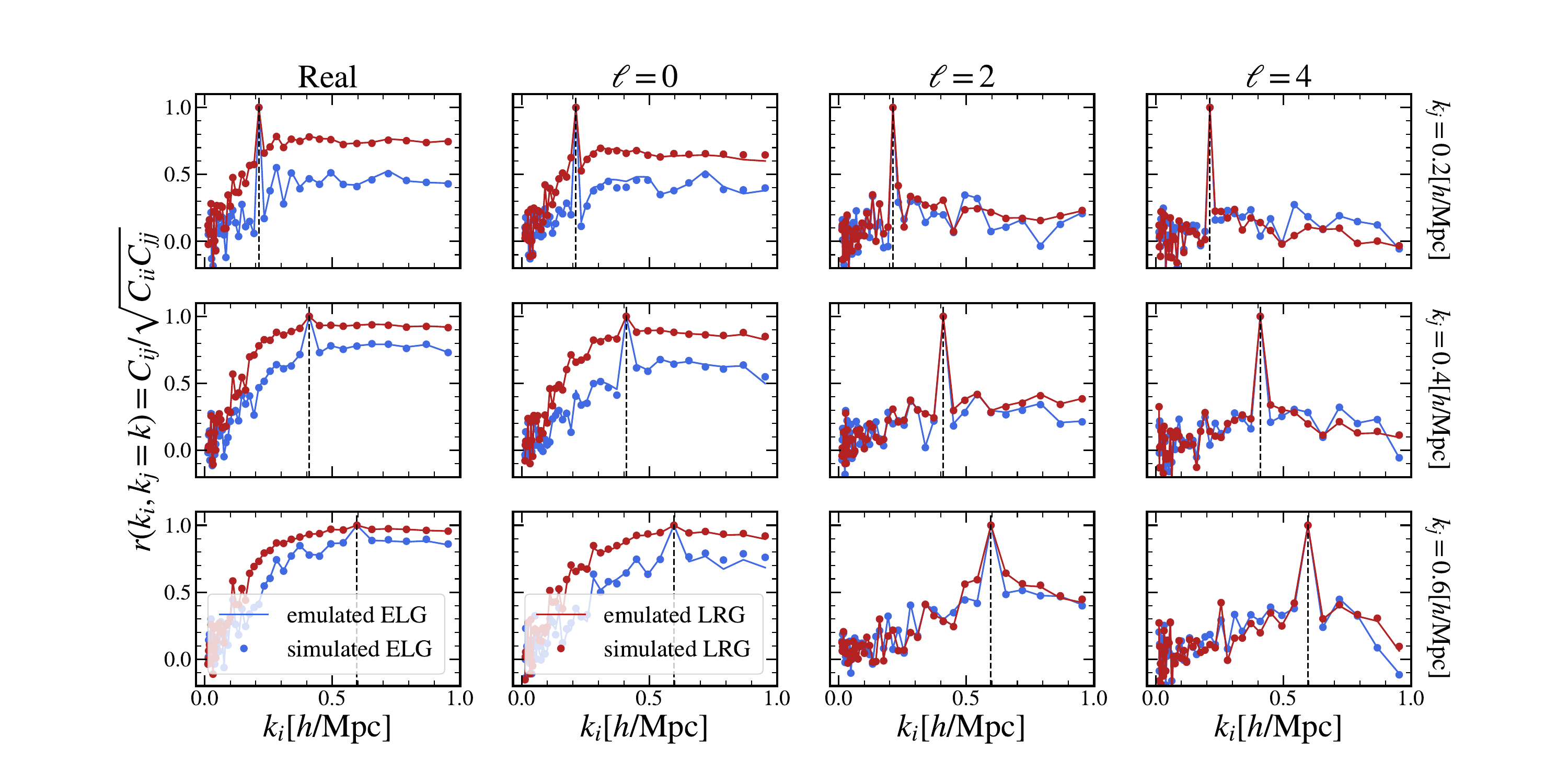}
    \caption{Upper Panels: the standard deviation predicted by \mtm before and after marginalization over the noise and FoG (Finger-of-God) nuisance parameters. The scatter, represented by points, corresponds to the Quijote standard deviation. Crosses indicate the Quijote standard deviation after applying FoG. The thick lines display the scatter recovered from \mtm results, while the dashed lines indicate the \mtm results after marginalization over the nuisance parameters. The different lines indicate the accuracy for distinct galaxy samples, namely LRG (Luminous Red Galaxies) and ELG (Emission Line Galaxies). Dark and light grey regions represents the 1\% and 3\% regions respectively. Lower Panels: the elements of the correlation coefficients matrix at a fixed $k_j$. These results are shown exclusively after marginalization over the nuisance parameters. The dashed line denotes $k_j=\{0.2,0.4,0.6\} \ihMpc$, which corresponds to the point of maximum correlation.} 
    \label{fig:redshiftspacegalvariance}
\end{figure*}

Combining the hybrid bias model with \mtm emulation significantly enhances the speed and efficiency of mock generation, enabling the estimation of large covariance matrices. Accurate covariance estimation is typically limited by the cost of generating the enormous number of realisations needed to reduce both the noise and the bias of our estimators (see e.g. \citealt{Dodelson2013}). This is crucial for obtaining rigorous constraints from large-scale structure observations. Our combined approach offers an ideal solution by enabling the creation of millions of accurate realisations at a moderate computational cost.

To validate our approach, we perform a comparison between the covariance matrix obtained from our suite of $100$ simulations and those obtained from the corresponding emulated galaxy fields. The complete structure of the correlation coefficients is illustrated in Fig.~\ref{fig:covariances}. These coefficients are defined as $r(k_i,k_j)=C_{ij}/\sqrt{C_{ii}C_{jj}}$, where $C_{ij}$ represents the covariance matrix obtained from 100 independent samples under the fiducial cosmology. The lower triangular sections of the matrices display the emulated results, whereas the upper triangular regions depict the simulated results. Upon visual inspection, the structures appear indistinguishable, suggesting that \mtm accurately reproduces the four-point functions as predicted by the hybrid bias model.

To make these comparisons more quantitative, the upper panels of Fig.~\ref{fig:redshiftspacegalvariance} depict these comparisons for both ELG and LRG samples diagonal terms, considering real space and the multipoles of the redshift-space power spectra. We present results for both the emulated model and the calibrated model, in which the nuisance parameters of the bias model have been refitted to accurately describe the emulated galaxy power spectra. It is worth noting that the parameters $\epsilon_1$ and $\epsilon_2$ have no impact on the covariance since they simply contribute constant values to the mean. Conversely, the redshift space nuisance parameters do influence the covariance as they enter through convolutions of the fields. Our findings reveal that the calibration process yields a slight enhancement in the monopole signal-to-noise ratio. However, it also leads to an apparent degradation in the variances of the quadrupole and hexadecapole spectra on intermediate scales. Although not shown here, the value of $\sigma/P$ seems, however, unaffected. Nevertheless, even with this degradation, the values remain within a 2\% range of accuracy.

In the lower panels of Fig.~\ref{fig:redshiftspacegalvariance}, we present the correlation coefficients $r(k_i,k_j)$ at $k_j=\{0.2,0.4,0.6\} \ihMpc$ for both real and redshift space outcomes. We find the biggest differences at the smallest scales, reaching values of 3-5\% at $k_i>0.7 \ihMpc$ in the monopole.

To put these numbers in context, \cite{Blot2019} performed a covariance matrix challenge based on the Minerva simulation suite \citep{Minerva}, where the typical accuracy reached 5-10\% down to scales of $k \sim 0.2 \ihMpc$. This methodology outperforms the methods explored in these previous work. However, more recent work, such as those of the \texttt{BAM} team claim subpercent results using similar ideas to machine learning approaches (see \citealt{Balaguera2019b, Balaguera2019a}, \citealt{Pellejero2020a}, \citealt{Sinigaglia2020}, \citealt{Kitaura2020}, and \citealt{Balaguera2023}). A comparison with such methods is beyond the scope of this work.

Note that 100 mocks are not enough to compute an unbiased covariance matrix. In an ideal case, the Hartlap factor $H$ \citep{Hartlap2007} gives us an estimation of the accuracy of our covariance given the number of mock catalogues. Concretely, for the number of data points and mocks explored here, $H\sim 0.53$, which means that our covariance estimation is off by this amount. However, both emulated and simulated results suffer the same effect since they share the same initial conditions of the simulations and the comparison performed here holds.

Summarising, we find that \mtm is indeed able to accurately recover the variance of the power spectra. The deviations from the $N$-body values are smaller than 3\% at scales $k_i>0.7 \ihMpc$ and typically at subpercent levels. There remain several key aspects that require further validation before using the approach introduced here for covariance estimation. For instance, the ability of the hybrid bias model to correctly reproduce the expected covariance (i.e. testing whether the trispectrum is well described, the role of higher-order bias parameters, etc), and extensions of \mtm to lightcones and higher redshifts need to be further developed and explored. We plan to carry out these investigations in future work.

\subsubsection{Cosmology dependence}

\begin{figure*}
    \includegraphics[width=\textwidth]{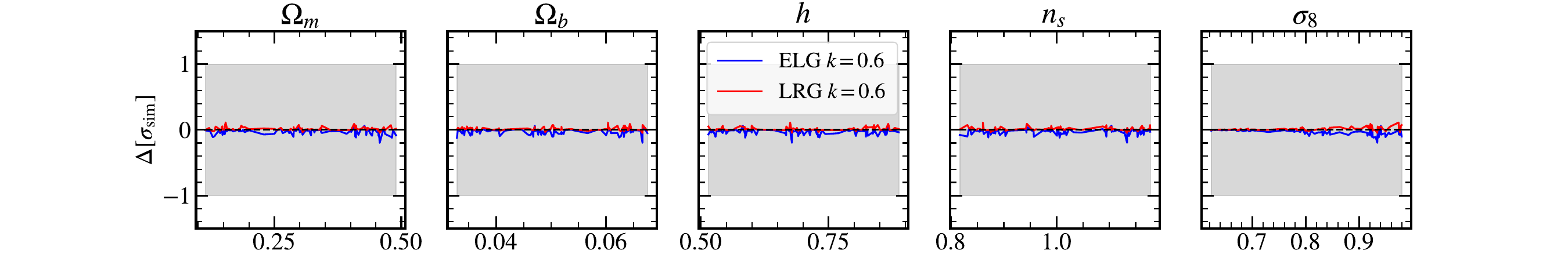}
    \includegraphics[width=\textwidth]{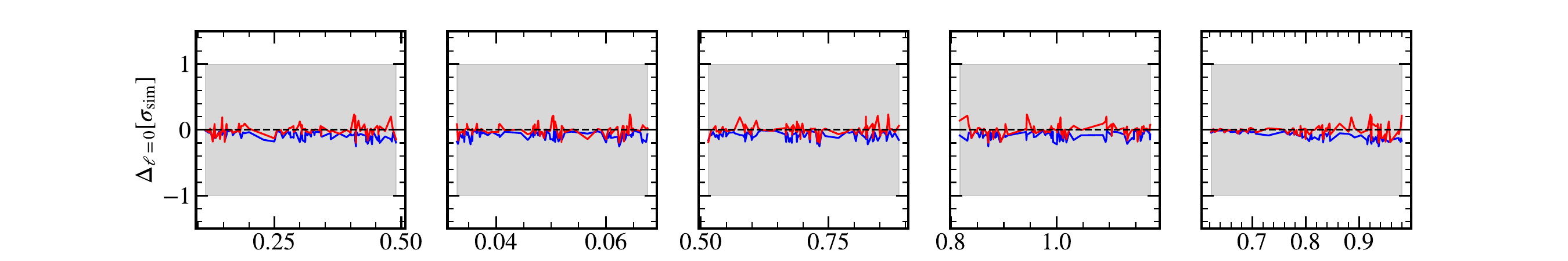}
    \includegraphics[width=\textwidth]{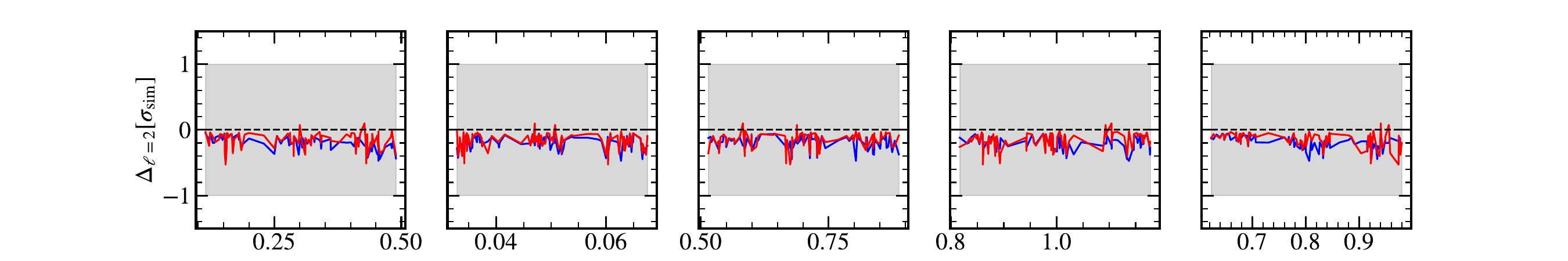}
    \includegraphics[width=\textwidth]{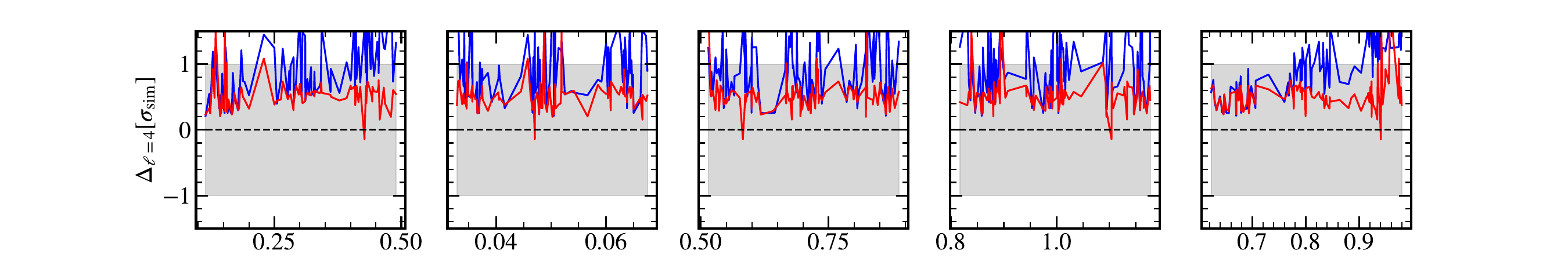}
    
    \caption{Difference of the power spectrum between the \mtm emulation and the galaxy simulated mock in units of the simulation standard deviation at a specific wavenumber ($k=0.6 \ihMpc$). Each column corresponds to a different cosmological parameter, while each row represents the deviation $\Delta[\sigma_{\rm sim}]\equiv (P_{\rm m2m} - P_{\rm Quijote})/\sigma_{\rm Quijote}$, which quantifies the difference in the 2-point statistics (in redshift or real space) normalised by the standard deviation of the Quijote simulations. Grey panels show the 1-$\sigma$ region.}
    \label{fig:errors}
\end{figure*}

As mentioned earlier, our combined approach shows great promise for analysing LSS and placing constraints on cosmological parameters. The effectiveness of incorporating \mtm into these analysis strategies depends on it reliably predicting the cosmological dependence of the 5 advected Lagrangian fields in our bias expansion.

We explore this issue in Fig.~\ref{fig:errors}, where we compare the \mtm predictions for ELG and LRG galaxy power spectra in 100 different cosmologies of the Quijote Latin Hypercube suite. Specifically, we display the difference in units of the fiducial standard deviation between these two estimates at the wavenumber $k=0.6\ihMpc$ as a function of the value of $\Omega_m$, $\Omega_b$, $h$, $n_s$, and $\sigma_8$. Note that we employ the calibrated versions of our parameters where the noise and RSD nuisance parameters are chosen so that they capture inaccuracies in \mtm. These parameters are varied to fit each cosmology independently, as it would be done in an analysis of observational data. We have verified that this scale is representative of the biggest differences at every cosmology.

The accuracy we achieve when varying cosmological parameters is almost entirely consistent with the results found earlier at the fixed, fiducial cosmology. The errors are within $1\sigma_{\rm sim}$ for the real and redshift space power spectra. The only exception is the hexadecapole, where the emulator's inherent scatter results in values exceeding $1\sigma_{\rm sim}$, particularly for the ELG sample. We have checked that this agreement extends to the range of scales shown in the bottom-most panel of Fig.~\ref{fig:redshiftspacegalpower}. This is a remarkable result as it demonstrates that \mtm is capable of generating biased fields at the level of scatter observed in $N$-body simulations across a broad range of cosmologies and scales.

We find the only significant cosmology dependence of the errors is on $\sigma_8$, which exhibits marginal trends that are consistent across different statistics. Specifically, for lower values of $\sigma_8$, we observe lower deviations and reduced scatter. Conversely, for typical values of $\sigma_8>0.8$, the scatter doubles, and the deviations become more pronounced. The most prominent example is the hexadecapole, which exhibits deviations of up to $2\sigma_{\rm sim}$ for $\sigma_8\sim 0.95$. This trend aligns with the findings of \cite{Jamieson2022}, although in their case, the lack of a calibration procedure exacerbates these effects. Fixing this dependency goes beyond the scope of this paper and we leave it for further work.

\subsection{Field-level}
\label{sec:field}

\begin{figure*}   
    \includegraphics[width=\columnwidth]{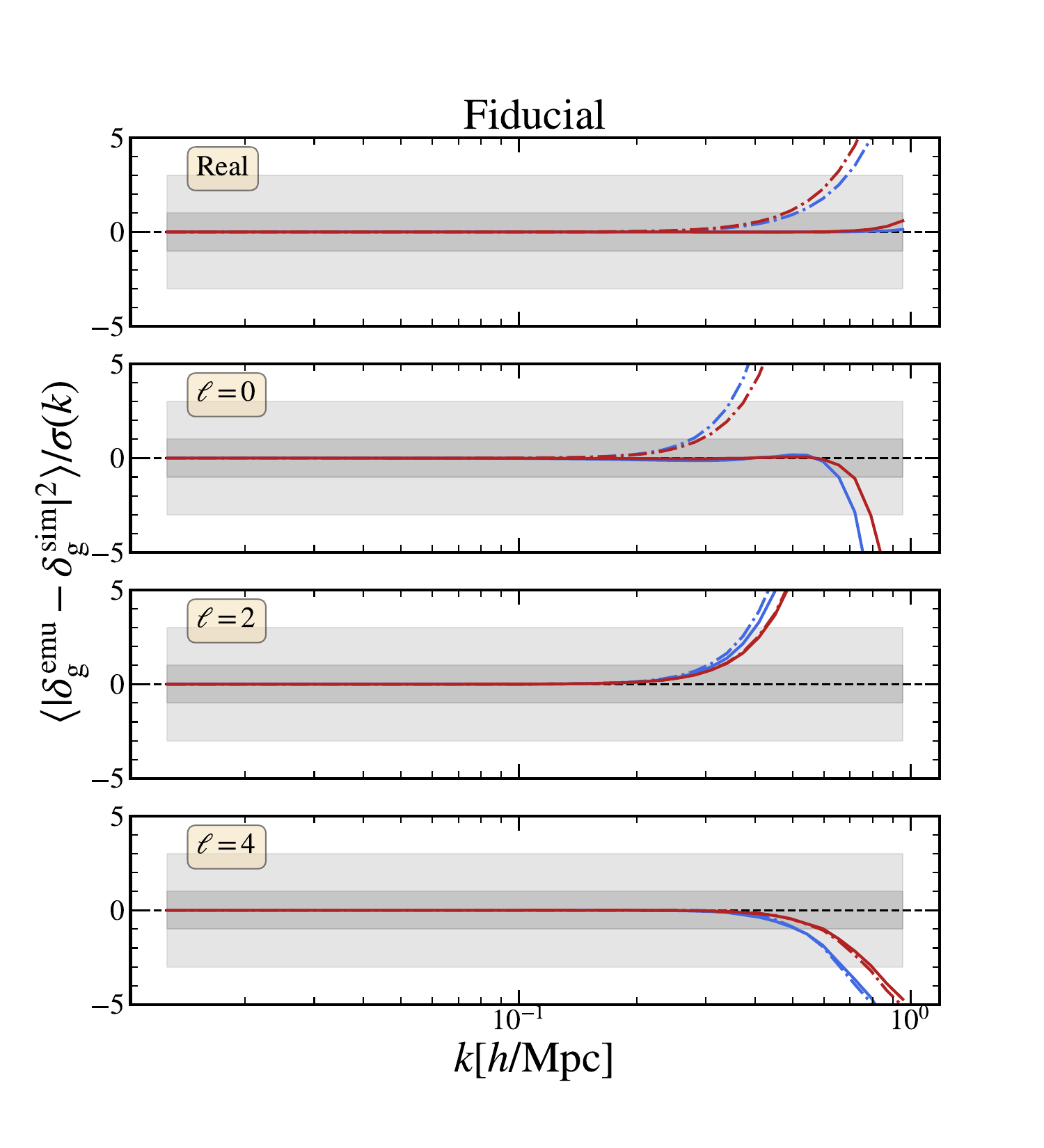}
    \includegraphics[width=\columnwidth]{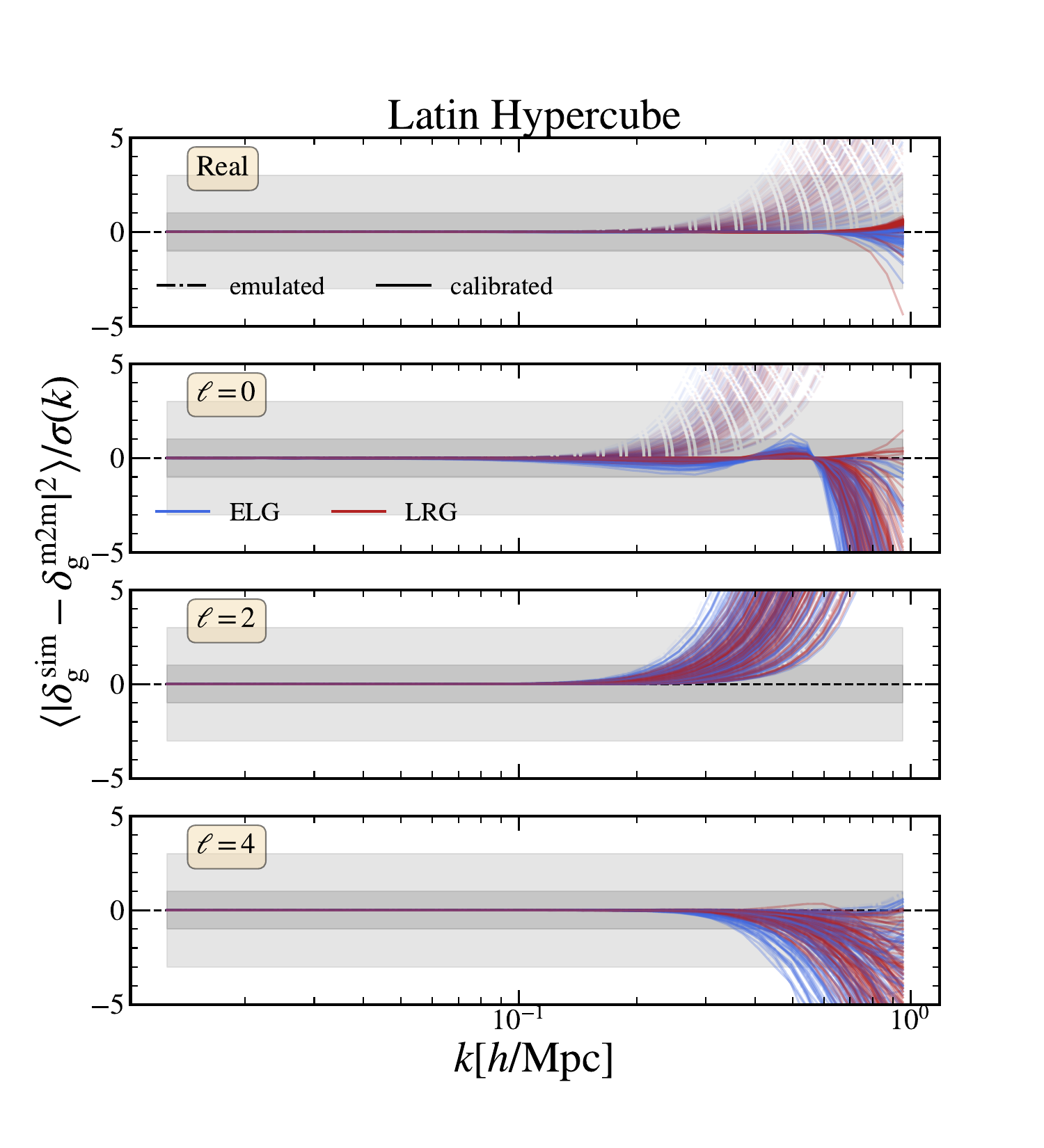}
    \caption{Expected difference in the amplitude of each Fourier mode between the LRG and ELG galaxy simulated fields as predicted by \mtm compared to the Quijote simulations weighted by the Gaussian variance at each mode. The presented panels include results for real space and redshift space multipoles of the field residuals. Dash-dotted lines indicate results from the emulation of the fields directly. Full lines represent results after calibration through the free nuisance parameters. Left Panels: The semi-transparent lines represent individual results from 100 Quijote simulations randomly selected from the Latin Hypercube (LH) space cosmologies. Right Panels: The thick coloured lines indicate the average results obtained from 100 Quijote simulations at the fiducial cosmology. Dark and light grey regions represents the 1-$\sigma$ and 3-$\sigma$ regions respectively.}
    \label{fig:galfieldlevel}
\end{figure*}

In previous sections we explored results for 2-point statistics, which facilitate interpretation and analyses that are familiar and easy to understand. However, the advantage of our model is that it provides predictions for the complete three-dimensional field, which allows us to go beyond 2-point statistics and consider the field-level information content. This has the potential to significantly increase the constraining power of extra-galactic survey observations \cite[e.g.][]{Porqueres2023}. Therefore, we now shift our focus to assessing the accuracy of \mtm at the field level.

\subsubsection{Fiducial cosmology}

As demonstrated in the bottom panel of Fig.~\ref{fig:example}, the comparison of the nonlinear density fields results in variations of approximately $\sim 5\%$ in the regions with the highest and lowest densities, generally staying lower. In the case of the velocity field, differences can exceed $\sim 20\%$ within haloes but decrease to less than $< 5\%$ outside of them. It's worth noting that this represents a significant improvement compared to the Zeroth Order Approximation (ZA) approach, where residuals exhibit values of 50\% consistently over extended regions, both inside and outside of haloes. 

The accuracy of field-level information, however, can be assessed using various methods (see e.g. \citealt{Cabass2020,Villaescusa-Navarro2021a}). Here, we adopt the approach presented in \cite{Schmittfull2019}, which computes the power spectrum of the difference between the simulated and emulated overdensity fields: $\Delta(k) \equiv \langle |\delta_{\rm g}^{\rm sim}-\delta_{\rm g}^{\rm m2m}|^2 \rangle$. Essentially, this corresponds to measuring the expected difference in the amplitude of each Fourier mode as predicted by the simulation and the emulation. The selection of this metric is based on its significance as the primary element featured in the field-level likelihood outlined in \cite{Kostic2023}. By opting for this specific metric, our aim is to examine the behaviour of the component of the field-level likelihood added across various scales. As with the case of the 2-point statistics, we will examine and compare two scenarios: one where we consider only deterministic contributions to the galaxy field, disregarding the nuisance parameters, and another where we incorporate their contributions. We compare these results against the expected Gaussian variance of each Fourier mode. Since on small scales we expect additional non-Gaussian contributions to the variance, our results will provide a conservative estimate to the accuracy of our hybrid approach. In redshift space, we show the multipoles of the resulting difference field.

The left panel of Figure \ref{fig:galfieldlevel} depicts the mean of $\Delta(k)$ computed over the 100 fiducial cosmology simulations, represented by red and blue lines. Dash-dotted lines represent the emulation results. The values of $\sigma(k)$ show the scatter within each $k$-bin, as estimated using the Gaussian approximation, $\sigma(k)=\sqrt{2/N_{\rm modes}}(P(k)+1/\bar{n})$ in real space, and the relation described in \cite{Grieb:2015bia} in redshift space. The figure indicates that real fields (upper panel) are recovered by \mtm with an error smaller than the expected contribution of discreteness noise down to scales of $\sim 0.5 \ihMpc$. In the second, third, and fourth panels, we present redshift space results. Emulated results begin to deviate at scales of around $\sim 0.3 \ihMpc$ for the monopole, $\sim 0.35 \ihMpc$ for the quadrupole, and $\sim 0.5 \ihMpc$ for the hexadecapole. Both ELG and LRG samples exhibit similar levels of accuracy. Notably, all results exhibit larger discrepancies compared to the outcomes derived from the power spectrum.

Thick lines show results after the absorption of emulator errors by the nuisance parameters of the model. Specifically, we implement this calibration by noting that we can expand $<|\delta^{\rm sim}-\delta^{\rm m2m}|^2> = <|\delta^{\rm sim}|^2> - 2<|\delta^{\rm sim}\delta^{\rm m2m}|> + <|\delta^{\rm m2m}|^2> + 1/n(\epsilon_1+\epsilon_2k^2) $, which yields the lowest values when minimising the field difference. For the redshift space case, the minimisation equations become more involved because of the inclusion of the factor $[(1-f_s)+f_s\lambda^2/(\lambda^2+k^2\mu^2)]$ in front of each $\delta_e$, but the quantity to minimise remains the same. The real space calibration yields unbiased results down to scales of $\sim 1 \ihMpc$. Calibrating nuisance parameters leads to improvement in the monopole, achieving an accuracy within 1$\sigma$ down to scales of around $\sim 0.6 \ihMpc$. On the other hand, the quadrupole and hexadecapole are largely unaffected by the calibration, unlike in the power spectrum case. This shows how much more stringent the field level emulation is. Note that the quadrupole and hexadecapole are mainly affected by the FoG parameters and are independent of the noise parameters. Therefore, it is possible that higher-order terms in the noise expansion may absorb the residual dependencies observed in this figure. We plan to explore these terms, which typically have $k^2\mu^2$ dependencies \citep{Perko2016}, in future studies.

Our results hints that \mtm can serve as a valuable tool for cosmological parameter inference at the field level. However, robust field-level parameter inference using the proposed hybrid model in combination with \mtm still requires two further validation tests. First, we must demonstrate that the results obtained from \mtm emulation hold when cosmological parameters are varied. Second, we need to investigate whether the hybrid model can deliver unbiased cosmological constraints at the field level. We focus on the first one next and leave the second for future work.

\subsubsection{Cosmology dependence}

We now check the accuracy of \mtm at the field level as a function of the 5-dimensional parameter space Latin Hypercube cosmologies of the Quijote suite \citep{QuijoteSims}. This approach enables us to assess the cosmological dependence of our earlier results. Similar to the fiducial case, we conduct the tests using the metric $\Delta(k)$.

The results of the cosmology dependence analysis are presented in the right panel of Fig.~\ref{fig:galfieldlevel}. The semi-transparent lines depict the root mean square (r.m.s) of each Fourier mode, weighted by their expected scatter, providing a measure of the field-level recovery when altering cosmological parameters. Our observations reveal that the results for the fiducial cosmology fall within the range obtained from the Latin Hypercube. However, this test exhibits a larger scatter compared to the power spectrum case. Deviations from the 1-$\sigma$ range in real space span scales from approximately $\sim 0.3 \ihMpc$ to $\sim 0.8 \ihMpc$. Similarly, for redshift space, deviations range from $[0.2,0.4]\ihMpc$ for the monopole, $[0.2,0.5]\ihMpc$ for the quadrupole, and $[0.4,1]\ihMpc$ for the hexadecapole.

Similar to the fiducial cosmology scenario, the thick lines in the results illustrate the outcomes after the absorption of emulator errors by the nuisance parameters of the model. Likewise, in real space, the nuisance parameters can absorb errors, extending the 1-$\sigma$ deviations to scales larger than $\sim 0.7 \ihMpc$. The monopole achieves the same level of accuracy down to scales of $\sim 0.6 \ihMpc$. Once more, higher multipoles are generally unaffected by the presence of nuisance parameters. This shows that the accuracy of \mtm, even when attempting to mitigate its inaccuracies through free parameters, remains dependent on the cosmological context. Although not explicitly presented here, the improvement in performance is correlated with a smaller value of $\sigma_8$, like the observations in the power spectrum case.

Remarkably, within the range of $\sigma_8$ values supported by current data (e.g., $\sigma_8 \sim 0.8$, as indicated by \citealt{CosmoPlanck18}), \mtm accurately reconstructs redshift space, surpassing expected noise levels down to scales of $\sim 0.35 \ihMpc$. This scale serves as a reference for studies focused on individual scales. However, for analyses such as cosmological parameter inference, where the emphasis is on the sum of differences (e.g., for computing $\chi^2$), using the mentioned scale may result in biased inferred cosmological parameters, as the sum of the $\Delta(k)$ differences adds up to 2-$\sigma$. A more conservative scale for such analyses would be $\sim 0.2 \ihMpc$, where the sum of $\Delta(k)$ adds up to less than 0.2-$\sigma$. In contrast, real space results suggest that \mtm can be applied across various scales down to $\sim 0.6 \ihMpc$. The consistent and reliable predictions across diverse cosmologies underscore the robustness and versatility of \mtm, presenting opportunities for its application in a wide range of field-level analyses.

\section{Conclusions}
\label{sec:conclusions}

In this paper, we have explored the idea of combining a field-level emulator with a hybrid bias expansion. Our emulator, \mtm, takes the 1LPT prediction as input and transforms it with a CNN model to predict the expected nonlinear displacement and velocity fields at $z=0$. In turn, the hybrid bias expansion is a general way to describe the abundance of galaxies. Combining the emulated nonlinear advection with the Lagrangian bias expansion, the hybrid bias expansion offers a general solution to model LSS tracers such as galaxies at the field level.

By comparing biased fields constructed using either simulations from the Quijote simulation suite or predictions from \mtm, we were able to quantify the accuracy of our approach for the power spectrum in real and redshift space, as well as for their variance, the field statistics, and cosmological dependence. 

We focused on two samples mimicking galaxies to be observed by the upcoming surveys -- Emission-line galaxies with relatively modest bias parameters at $z\sim1$ and Luminous red galaxies with larger bias parameters at somewhat lower redshifts $z\sim0.6$. 

In real space, we find that the galaxy fields predicted by our approach closely describe our simulation results on large scales. On intermediate scales, although accurate at the 1\% level, there are statistically significant biases. These, however, can be almost completely absorbed by the nuisance parameters of the hybrid bias expansion that describe the stochastic component of galaxy bias. 

Similarly, in redshift space, we found that our emulated results are accurate on large scales but are systematically biased for $k>0.2\,\ihMpc$ (in the quadrupole). We attribute this to an additional small-scale velocity dispersion in the \mtm predictions. This likely is the result of inaccuracies in exactly detecting the boundary of haloes. However, these too can be absorbed by the nuisance parameters of the redshift-space modelling. Specifically, by the parameter that describes the intra-halo velocity dispersion of the target galaxy samples, which leads to accuracies of within $1\sigma$ down to scales of $k\sim0.6\,\ihMpc$.

Additionally, we showed that our approach can accurately predict the variance and off-diagonal covariance terms of real and redshift space quantities. This opens up the possibility of estimating covariance matrices with very little numerical noise as our approach can deliver thousands of mock catalogues at a small computational cost.

Through tests of field-level performance, we have observed that \mtm accurately reconstructs both real and redshift space fields, surpassing the anticipated noise levels present in the considered mock datasets. In particular, the accuracy extends to scales of approximately $k \sim 1 \ihMpc$ in real space and $k \sim 0.35 \ihMpc$ in redshift space, where the main limitation is the emulated quadrupole.

Remarkably, these outcomes persist even when varying the cosmological parameters, emphasising the suitability of our approach for precise parameter estimation at the field level. However, it is worth noting a dependency on $\sigma_8$, where larger deviations are observed at higher values of this parameter. We plan to address this issue in future works.

\section*{Acknowledgements}
MPI and RA would like to thank Francisco Maion for useful discussions. The authors acknowledge the support of the ERC-StG number 716151 (BACCO). MPI acknowledges the support of the “Juan de la Cierva Formaci\'on” fellowship (FJC2019-040814-I). REA acknowledges the support of the Project of excellence Prometeo/2020/085 from the Conselleria d'Innovació, Universitats, Ciéncia i Societat Digital de la Generalitat Valenciana, and of the project PID2021-128338NB-I00 from the Spanish Ministry of Science. This research was supported by the Munich Institute for Astro-, Particle and BioPhysics (MIAPbP) which is funded by the Deutsche Forschungsgemeinschaft (DFG, German Research Foundation) under Germany's Excellence Strategy – EXC-2094 – 390783311.
 
\section*{Data Availability}

The data used throughout this paper is available in the following sources. The Quijote simulations are publicly available at the Quijote webpage\footnote{https://quijote-simulations.readthedocs.io/en/latest/}. The hybrid model can be computed from the baccoemu code\footnote{https://baccoemu.readthedocs.io/en/latest/}. The \mtm code with the trained weights can be downloaded from github\footnote{https://github.com/dsjamieson/map2map\_emu/}. The rest of the products will be 
shared upon reasonable request to the authors.



\bibliographystyle{mnras}
\bibliography{example}


\appendix

\section{Redshift-space operators}
\label{App:errors}

\begin{figure*}
\includegraphics[width=\textwidth]{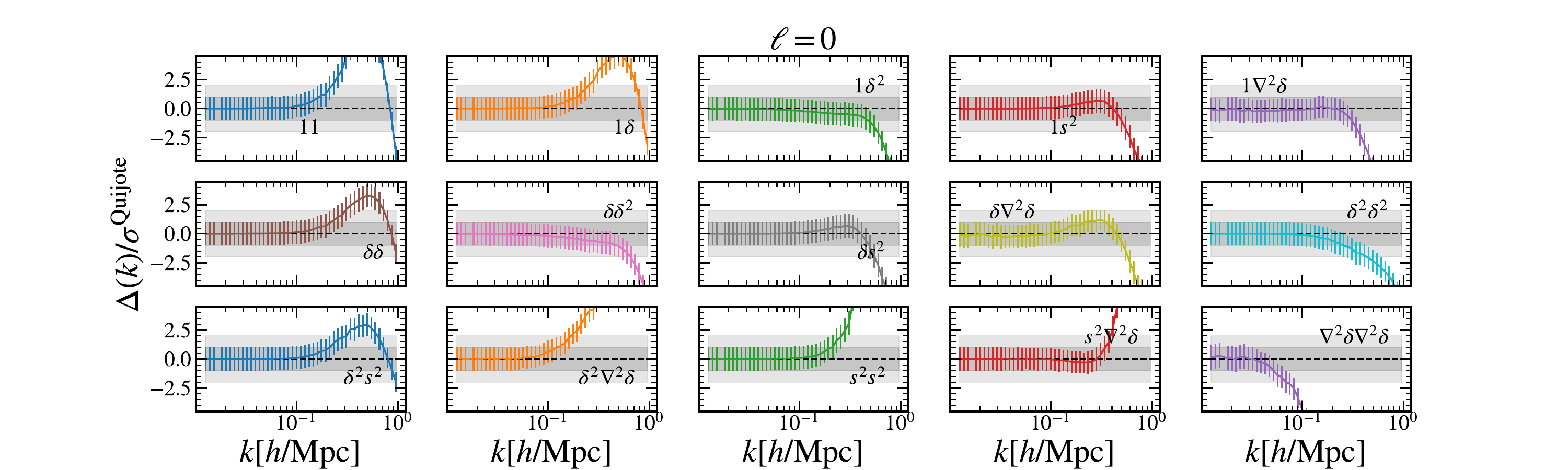}
\includegraphics[width=\textwidth]{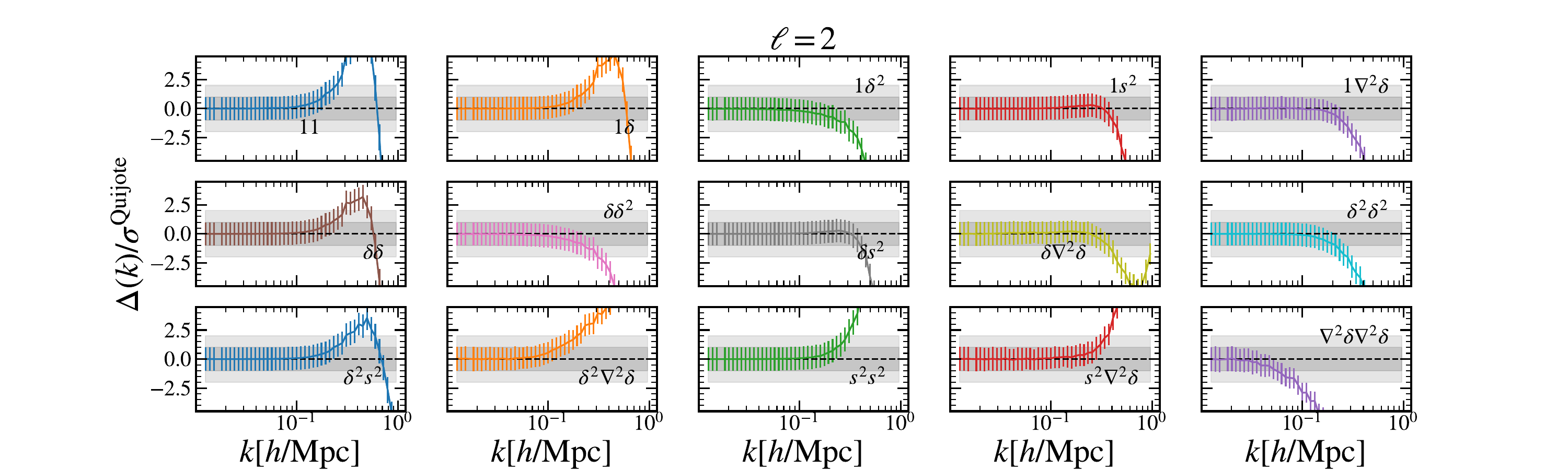}
\includegraphics[width=\textwidth]{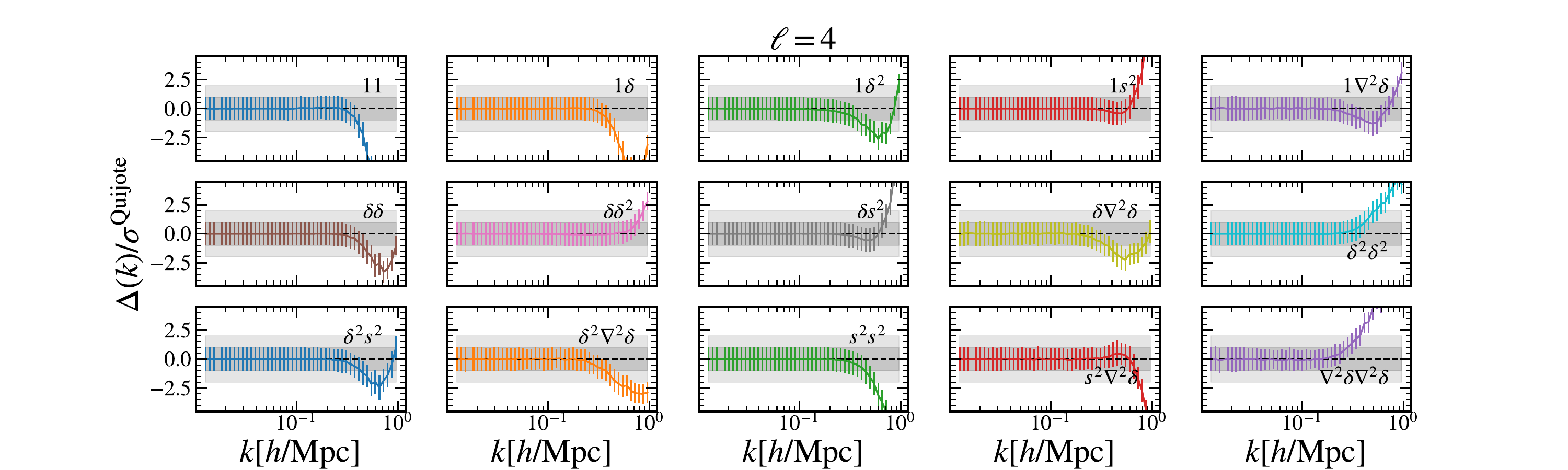}
\caption{The figure presents the relative difference between the predictions of \mtm and the measurements obtained from the Quijote suite, expressed in units of the simulation variance for a $1\hGpcC$ volume. Specifically, we consider the quantity $(P^\ell_{\rm m2m} - P^\ell_{\rm Quijote})/\sigma^\ell_{\rm Quijote}$, where $P^\ell_{\rm m2m}$ and $P^\ell_{\rm Quijote}$ represent the monopole, quadrupole, or hexadecapole power spectra in redshift space for each respective operator field advected to $z=0$. Each panel of the figure corresponds to a different operator field, as indicated in the respective captions. The solid lines represent the average relative difference, while the vertical error bars represent the standard deviation, calculated over 100 realizations.}
\label{fig:RSD_rel_diff}
\end{figure*}

The accuracy of the \mtm operators in redshift space is shown in Figure \ref{fig:RSD_rel_diff}, illustrating the fractional difference observed among each of the cross-spectra. To provide clarity, the differences are presented in terms of the standard deviation measured in the Quijote realisations. The error bars represent the scatter determined from the \mtm recovered spectra. Thus, a scatter closer to 1 indicates a better recovery of the diagonal terms of the covariance matrix. We explore this further in the main text.

On large scales ($k < 0.1 \ihMpc$), the emulated and simulated cross-spectra exhibit statistical compatibility. However, on smaller scales, systematic deviations are detected regardless of the multipole, with \mtm either over- or under-predicting the signal by more than $3\sigma$, depending on the specific spectra. It is worth noting that despite the presence of this bias, its magnitude remains extremely small. For example, for the $P_{11}$ term, the bias corresponds to only 1.5\% at $k=1\ihMpc$. The overall significance and relative importance of uncertainties in each $P_{i,j}$ term depend on the specific bias parameters characterising the given galaxy sample, which is explored in this work. Our findings reveal that, with realistic bias values, these differences can be effectively absorbed through the calibration of the model nuisance parameters, leading to results within $1\sigma$ for both the ELG and LRG samples.

\bsp	
\label{lastpage}
\end{document}